\documentclass[12pt]{spieman}  
\usepackage{amsmath,amsfonts,amssymb}
\usepackage{graphicx}
\usepackage{setspace}
\usepackage{tocloft}

\def\farcm{\hbox{.\kern -0.7ex\raisebox{.9ex}{\scriptsize$\prime$}}}
\def\farcs{\hbox{\kern 0.13ex.\kern -0.95ex%
\raisebox{.9ex}{\scriptsize$\prime\prime$}\kern -0.1ex}}

\def\lesssim{\mathrel{\hbox{\rlap{\hbox{\lower4pt\hbox{$\sim$}}}\hbox{$<$}}}}
\def\gtrsim{\mathrel{\hbox{\rlap{\hbox{\lower4pt\hbox{$\sim$}}}\hbox{$>$}}}}

\def\farcm{\hbox{.\kern -0.7ex\raisebox{.9ex}{\scriptsize$\prime$}}}
\def\farcs{\hbox{\kern 0.13ex.\kern -0.95ex%
\raisebox{.9ex}{\scriptsize$\prime\prime$}\kern -0.1ex}}

\def\la{\mathrel{\hbox{\rlap{\hbox{\lower4pt\hbox{$\sim$}}}\hbox{$<$}}}}
\def\ga{\mathrel{\hbox{\rlap{\hbox{\lower4pt\hbox{$\sim$}}}\hbox{$>$}}}}

\def\farcm{\hbox{$.\mkern-4mu^\prime$}}
\def\farcs{\hbox{$.\!\!^{\prime\prime}$}}

\def\fnum@figure{{\rmfamily Fig.\space\thefigure.---}}%
\def\fnum@table{{\rmfamily Table \thetable:}}%
\def\fnum@plate{{\bfseries Plate \theplate.}}%
\def\fps@figure{bp}%
\def\fps@table{bp}%
\def\fps@plate{bp}%

\title{Ground Layer Adaptive Optics for the W. M. Keck Observatory: Feasibility Study}

\author[a]{J. R. Lu}
\author[b]{M. Chun}
\author[c]{S.~M.~Ammons}
\author[d]{K. Bundy}
\author[e]{R. Dekany}
\author[f]{T. Do}
\author[d]{D. Gavel}
\author[h]{M. Kassis}
\author[i]{O. Lai}
\author[j]{C. L. Martin}
\author[d,g]{C. Max}
\author[k]{C. Steidel}
\author[l]{L. Wang}
\author[g]{K. Westfall}
\author[h]{P. Wizinowich}
\affil[a]{Astronomy Department, University of California, Berkeley, CA 94720, USA}
\affil[b]{Institute for Astronomy, University of Hawaii, Manoa}
\affil[c]{Lawrence Livermore National Labs}
\affil[d]{University of California Observatories}
\affil[e]{Caltech Optical Observatories, California Institute of Technology}
\affil[f]{University of California, Los Angeles}
\affil[g]{University of California, Santa Cruz}
\affil[h]{W.~M. Keck Observatory}
\affil[i]{Observatoire de la C\^{o}te d'Azur, CNRS}
\affil[j]{University of California, Santa Barbara}
\affil[k]{California Institute of Technology}
\affil[l]{Thirty Meter Telescope}

\cftpagenumbersoff{figure}
\cftpagenumbersoff{table} 
\begin{document} 
\maketitle

\begin{abstract}
Ground-layer adaptive optics (GLAO) systems offer the possibility of improving the ''seeing'' of large ground-based telescopes and increasing the efficiency and sensitivity of observations over a wide field-of-view. We explore the utility and feasibility of deploying a GLAO system at the W.~M.~Keck Observatory in order to feed existing and future multi-object spectrographs and wide-field imagers. We also briefly summarize science cases spanning exoplanets to high-redshift galaxy evolution that would benefit from a Keck GLAO system. Initial simulations indicate that a Keck GLAO system would deliver a 1.5$\times$ and 2$\times$ improvement in FWHM at optical (500 nm) and infrared (1.5 $\mu$m), respectively.  The infrared instrument, MOSFIRE, is ideally suited for a Keck GLAO feed in that it has excellent image quality and is on the telescope's optical axis. However, it lacks an atmospheric dispersion compensator, which would limit the minimum usable slit size for long-exposure science cases. Similarly, while LRIS and DEIMOS may be able to accept a GLAO feed based on their internal image quality, they lack either an atmospheric dispersion compensator (DEIMOS) or flexure compensation (LRIS) to utilize narrower slits matched to the GLAO image quality. However, some science cases needing shorter exposures may still benefit from Keck GLAO and we will investigate the possibility of installing an ADC.
\end{abstract}


{\noindent \footnotesize\textbf{*}Jessica R. Lu,  \linkable{jlu.astro@berkeley.edu} }

\section{Introduction}
Ground-layer adaptive optics (GLAO) provides partial correction of atmospheric blurring over a significantly larger field of view (several to 10 arcminutes) and over a broader wavelength range (optical and infrared) than classical AO systems \cite{Rigaut:2002}.  Science applications enabled by GLAO are thus complementary to classical AO and include extra-galactic spectroscopic surveys over a broad range of redshifts, intergalactic and circumgalactic medium studies with integral field and slit spectrographs, and stellar population studies in the Milky Way and nearby galaxies where crowding limits current sensitivities. Maunakea is ideally suited for GLAO with its thin ground layer and weak free-atmosphere turbulence \cite{Chun:2009}. Currently, none of the 8-10 m class telescopes on Maunakea have a GLAO system; however, several observatories are investigating the feasibility and designing GLAO systems for the future. In this proceeding, we explore the feasibility of a GLAO system deployed at the W. M. Keck Observatory, which could have the benefit of feeding Keck's existing seeing-limited instruments including the multi-object spectrographs, LRIS and DEIMOS in the optical and MOSFIRE in the infrared. 

Nearly all seeing-limited science cases would benefit from the expected 2$\times$ better spatial resolution in the near-infrared and red-optical delivered by GLAO \cite{Abdurrahman:2018,Chun:2018}.  The ultimate question that must be addressed is whether the science gains enabled by GLAO system at Keck are timely and significant enough to warrant the cost. The science gains delivered by a Keck GLAO system can be quantified with the following ingredients. First, we need to estimate the image quality delivered by a notional GLAO system. We utilize multiple atmospheric modeling codes and mean Maunakea atmospheric conditions to simulate point spread functions (PSFs) for Keck GLAO. We also use results from the on-sky 'imaka experiment to validate these simulations. Second, we need to determine how the existing instruments would further reduce the GLAO-delivered image quality due to internal aberrations, distortions, and flexure. Thirdly, we need to propagate these image quality estimates through to science metrics for a wide range of current and future science cases. Finally, science gains must always be weighed against cost and risk. In this proceeding, we present initial results from the first 2$-$3 steps of this process.  

\section{GLAO Performance: On-sky Validation and Keck Simulations}
One of the primary motivations for GLAO at Keck has been results from on-sky experiments that indicated that the turbulence above Maunakea is typically dominated by a significant, but thin, ground layer \cite{Chun:2009}. In particular, the 'imaka GLAO experiment commissioned on-sky on the UH 2.2 m telescope on Maunakea \cite{Chun:2016} has shown that the FWHM can be improved by a factor of 1.5$-$1.7$\times$ and the Noise Equivalent Area is improved by $>$2$\times$ at R-band and I-band with a guide star constellation spread over 18' \cite{Abdurrahman:2018}. Preliminary results from 2018 indicate that improvements of $>$1.3$\times$ are achievable even at B-band and V-band over an 11' field of view \cite{Chun:2018}. The GLAO PSFs are also more stable by a factor of 7-10$\times$.  The results of 'imaka are promising for a Keck GLAO system. 
Furthermore, GLAO performance has been validated at the VLT, showing 2$\times$ improved and very uniform FWHM over a 10' radius \cite{Madec:2018}. 

\begin{figure}
\begin{center}
\includegraphics[height=8cm]{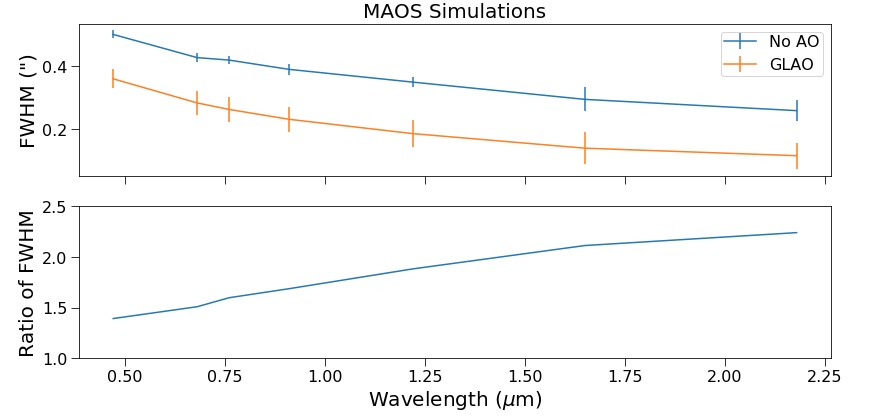}
\end{center}
\caption{ \label{fig:glao_sims}
Simulated image quality, expressed in FWHM, for Keck in seeing-limited and GLAO modes at different wavelengths ({\em top}). The ratio of the GLAO FWHM to the seeing-limited FWHM is also shown ({\em bottom}).
} 
\end{figure}

Simulations of the expected performance of the a Keck GLAO system were made using the Multi-threaded Adaptive Optics Simulator (MAOS) \cite{Wang:2012}.  The initial system configuration is for a 20$\times$20 actuator system with four sodium laser guide stars placed on the periphery of a 10'$\times$10' field of view.  A single low-order natural-guide star wavefront sensor was placed at the center of the field to provide tip, tilt, and focus sensing.  Point spread functions were generated across the 10' field for wavelengths from 0.45 to 2.2 $\mu$m from a sequence of ten independent 5000-iteration simulations.  The input optical turbulence profile is derived from the Thirty Meter Telescope project's site testing at the TMT site on Maunakea \cite{Schock:2009}. The results shown in Fig.~\ref{fig:glao_sims} were run to provide an initial estimate of the gains across the science wavelengths of interest and represent averages across the field of view.  One-standard deviation in this variation are indicated on the plot.  The improvement in the FWHM is 2$\times$ at infrared wavelengths and $\sim$1.5$\times$ at optical wavelengths ($\sim$500 nm).  Analysis of other metrics, such as the encircled energy and the noise-equivalent area, is underway.  Future simulations will focus on optimizing the image quality for the specific science instrument fields of view and wavelengths and including the effect of the conjugation altitude the adaptive secondary.  

\section{Examples of GLAO Science Cases}

In this section, we present a brief summary of science cases that would most benefit from GLAO fed multi-object spectroscopy at Keck. More detailed science cases for GLAO at other telescopes have previously been published for Subaru \cite{SubaruGLAO}, Gemini \cite{GeminiGLAOstudy,GeminiGLAOsci}, CTIO \cite{Tokovinin:2015}, and VLT \cite{Hibon:2016,Siebenmorgen:2011,Bacon:2006}.

Measurements of galaxy properties test our physical understanding of galaxy formation and evolution. 
The sensitivity of ground-based extra-galactic spectroscopy is limited by the sky brightness.  The improved spatial resolution delivered by GLAO is sufficient to 
resolve galaxies that were previously unresolved with seeing-limited observations. As a result, spectroscopic slits can be narrower, thereby reducing the solid angle of blank sky that adds noise to the galaxy spectrum.
Cutting the spectroscopic aperture down to the size of resolved star-forming regions within galaxies
also increases spectral resolution, which improves sensitivity to spectral lines. It follows that GLAO
enables spectroscopic observations of fainter galaxies.

\begin{figure}
\begin{center}
\includegraphics[height=8cm]{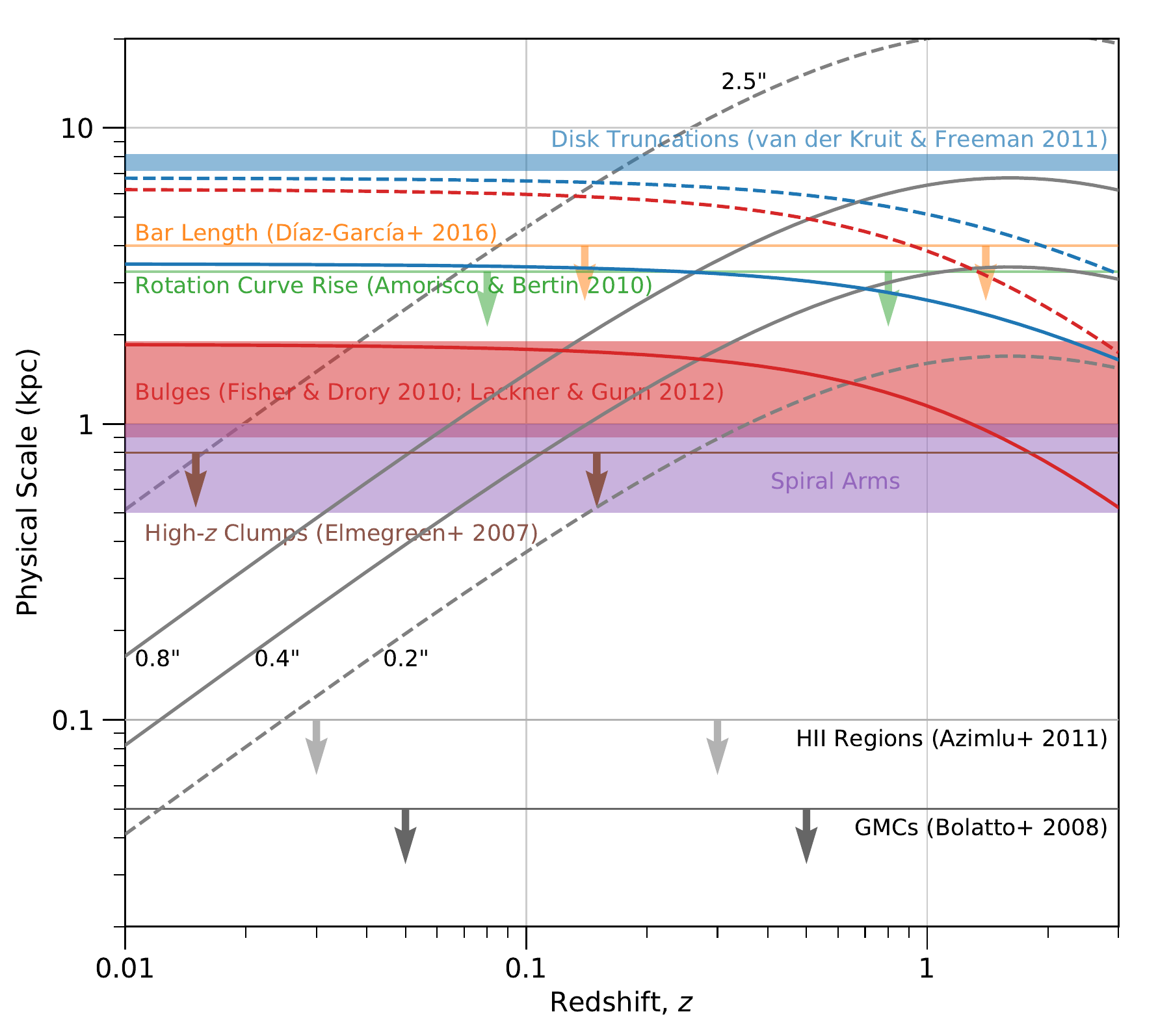}
\end{center}
\caption{ \label{fig:exgal_scale}
Interesting physical spatial scales in galaxies (e.g. disks, bars, bulges, clumps) as a function of redshift. Lines of constant angular resolution on the sky are over-plotted. GLAO at Keck could enable new spatially-resolved surveys of galaxies 3-5$\times$ higher redshift than seeing-limited surveys.
} 
\end{figure}

One of the great advantages of GLAO is how the improved spatial resolution allows us to resolve finer detail in distant galaxies (Fig.~\ref{fig:exgal_scale}). The power of spatially-resolved spectroscopy in large galaxy surveys has recently been demonstrated in the MaNGA survey using seeing-limited, integral-field, multi-object spectroscopy. A Keck GLAO system can push to higher redshift and lower-mass galaxies than current IFU surveys such as MaNGA. 

There are many additional GLAO science cases including:
(1) detailed studies of galactic nuclei,
(2) resolving stellar populations and star formation histories in nearby galaxies,
(3) mapping and characterizing the interstellar medium, supernovae remnants, and outflows from stars in the Milky Way,
(4) detection of exoplanet from the astrometric wobble of their host star,
(5) complete imaging of Mars, Jupiter, and other planets simultaneously with their ring and moon systems.
Most of these science cases utilize the improved spatial resolution of GLAO to resolve further detail on an object or overcome stellar confusion rather than the improved sensitivity.

\section{Instrument Feasibility}

\begin{figure}
\begin{center}
\includegraphics[height=8cm]{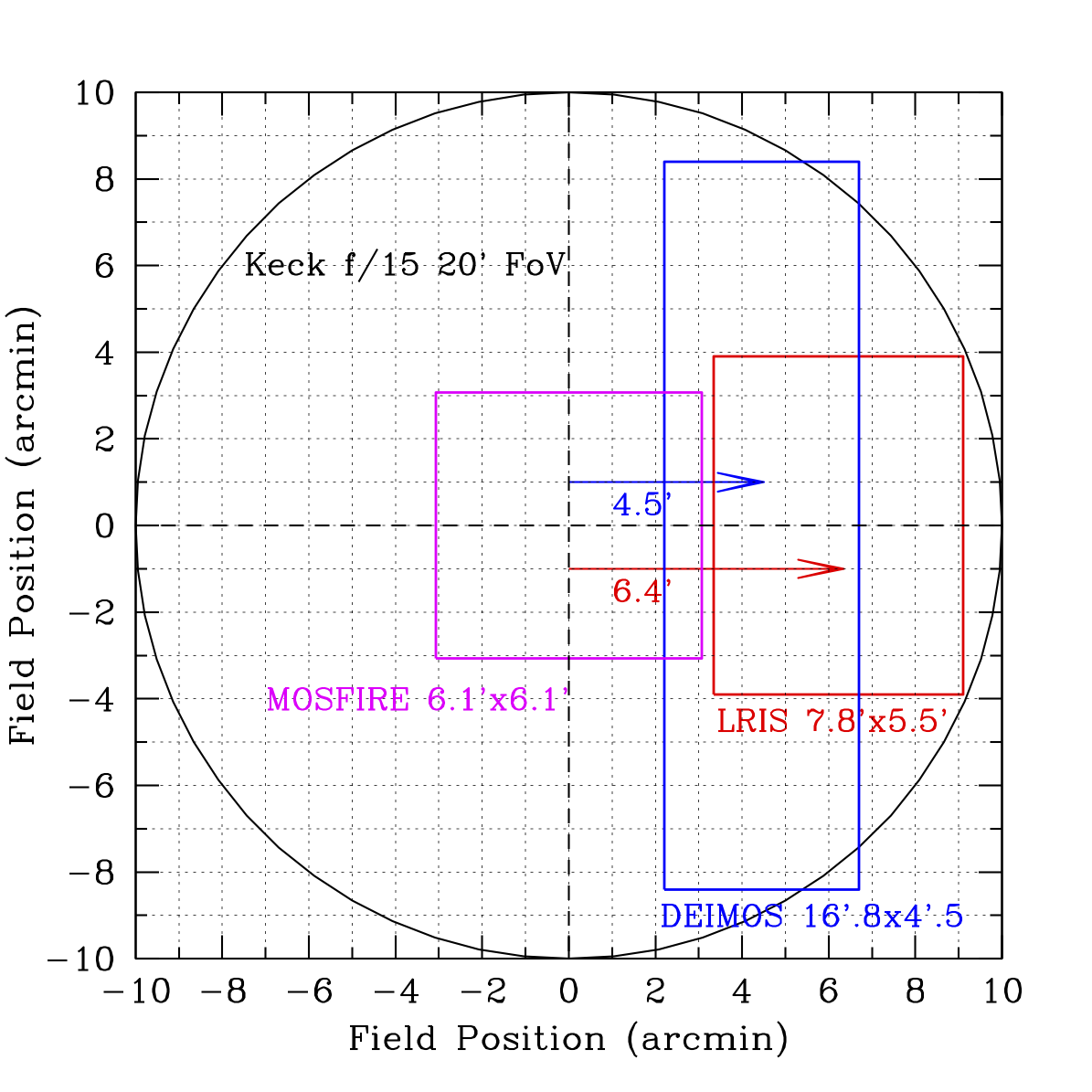}
\end{center}
\caption{\label{fig:instr_fov}
The Keck instrument fields-of-view relative to the optical axis of the telescope. Both LRIS and DEIMOS fields extend out to 10’ from off the optical axis and their image quality would likely be limited by off-axis telescope aberrations in the outer parts of the field. MOSFIRE is centered on the telescope optical axis.  
}
\end{figure}

Many of Keck's existing multiplexed spectrographs and wide-field imagers could benefit from a GLAO feed.  However, the feasibility of using existing instruments depends on many factors. First and foremost is whether the intrinsic optical quality of the telescope and instrument are sufficiently good to take advantage of the higher spatial resolution that GLAO would deliver. Figure \ref{fig:instr_fov} shows the locations and field sizes for different instruments relative to the optical axis of the telescope. We also consider other factors that might limit the science gains from a GLAO feed such as atmospheric refraction, flexure, and slit construction. These turn out to be the limiting factors in most cases for existing instruments. In sections below, we consider the feasibility of the two multi-object optical spectrographs, LRIS and DEIMOS, as well as the infrared multi-object spectrograph, MOSFIRE. 

\subsection{LRIS}

The multi-object, optical slit spectrograph, LRIS, is located on the optical axis of the telescope (Figure \ref{fig:instr_fov}) and thus has access to the best image quality directly from the telescope. It's 6' $\times$ 6' FoV, 0.135"/pixel plate scale in the imaging direction, and designed image FWHM of $\lesssim 0.24$''  should make it an ideal GLAO instrument \cite{Oke:1995}. A GLAO feed for LRIS would allow the use of smaller slits for higher spectral resolution and lower background. Typical slits in seeing-limited mode are 0.8'' - 1.0'' and we anticipate a slit width reduction to 0.5''. 

\begin{figure}
\begin{center}
\includegraphics[scale=0.4]{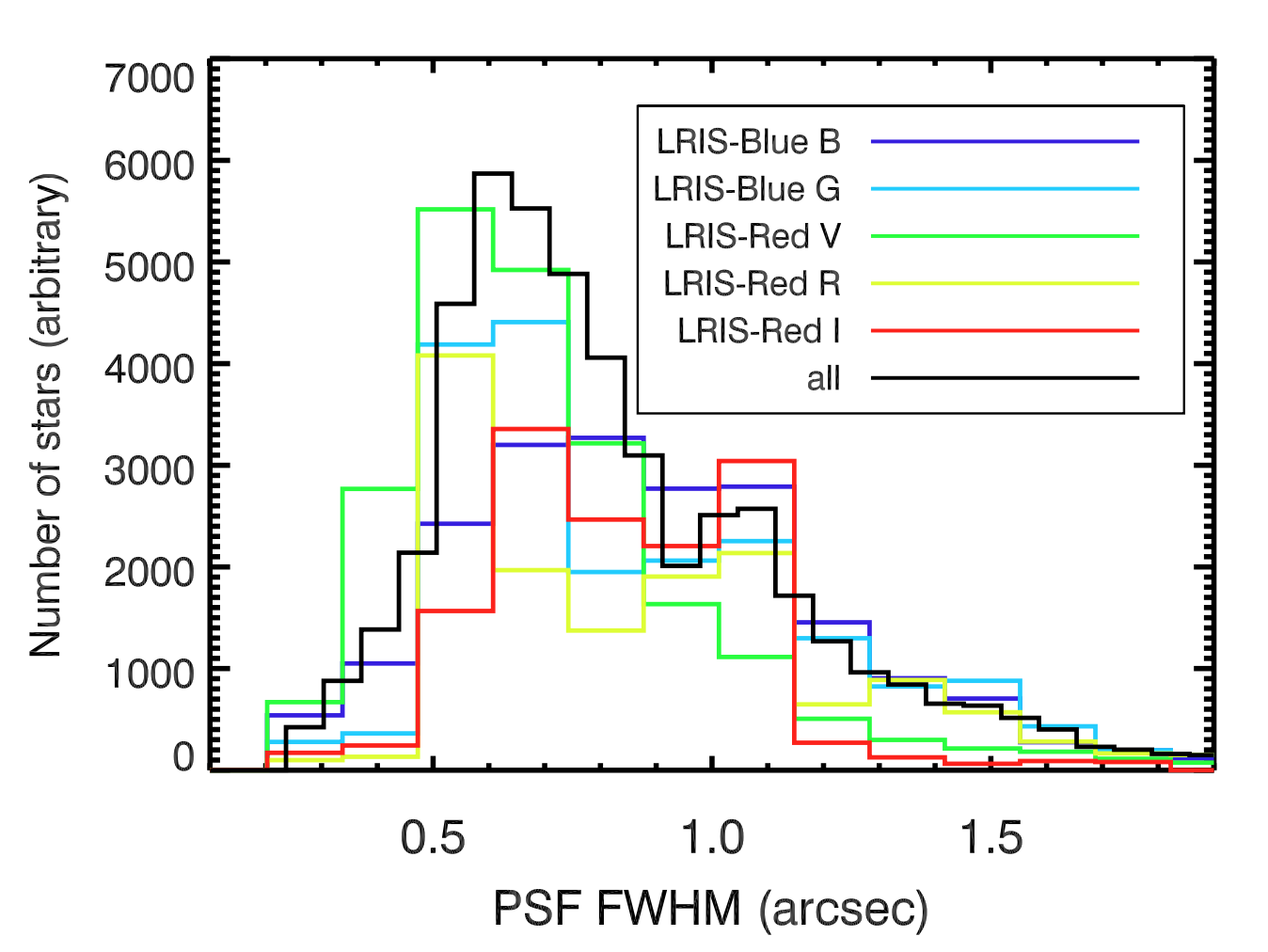}
\end{center}
\caption{The distribution of measured FWHM from LRIS images of dense stellar fields. Measurements come from a random sample of 126 images obtained after 2011 Jan. 
\label{fig:lris_fwhm}}
\end{figure}

First, the as-delivered image quality on LRIS has been verified with on-sky images from both the LRIS Blue and Red camera. We collected a random sample of $\sim 400$ LRIS images from observations taken after 2011 Jan from the Keck Observatory Archive. Stars were identified using DAOPHOT with a number of criteria to ensure that resolved sources and artifacts were excluded. PSF fitting was performed on each star and the median stellar PSF FWHM for the data set was calculated (Figure \ref{fig:lris_fwhm}).  The median image FWHM for all cameras and filters is 0.836''.  The discrepancy between that value and the expected value of 0.67'' (atmospheric free seeing of 0.65'' and contribution from telescope astigmatism of 0.18'') suggests the presence of 0.49'' of unaccounted-for wavefront error during typical LRIS usage. The telescope contributes $\sim$0.18'' averaged over the field due to off-axis field aberrations. Thus, an unexplained term of 0.5'' remains and is likely due to slowly time-varying focus or astigmatism and should be correctable by the GLAO system.

Using realistic GLAO PSFs with the appropriate ASM conjugate altitude, we calculate exposure time improvement factors of 10\% to 140\% (i.e., t/1.1 $-$ t/2.4) for stars within a 4' $\times$ 4' field at 8000 $\AA$.  We assumed the GLAO system parameters include 3-guide stars with a constellation diameter of 4'.  These improvement factors depend on a number of parameters, including the line width, assumptions about the distribution of correctable and non-common path wavefront error, and other factors not explored here.  The most important factor is the contribution of non-common path wavefront error.  The median image quality obtained with LRIS suggests that this factor could be as high as 0.49''.  If it is, the improvement in required exposure time seen with GLAO would be small.  However, if the error is common-path and could be sensed and corrected by the GLAO system, the improvement in required exposure time with GLAO could be as high as 150\%. 

In addition to the image quality, we consider other sources of aberration and distortion. LRIS has an atmospheric dispersion compensator, so differential atmospheric refraction should not contribute significantly for a GLAO PSF. On the other hand, LRIS is a Cassegrain instrument and is know to suffer from time-variable distortion and focus. Previous studies of the astrometric distortion \cite{Cohen:2007} indicate that in seeing limited mode, the flexure amounts to 0.1'' - 0.2'' over the field. However, the same document states that flexure of more than 10\% of a slit width is not tolerable. Thus if the LRIS slit sizes are reduced from 0.8'' to 0.5'' then the LRIS flexure would not keep the science object on its slit. These flexures need to be more extensively quantified in order to determine if some real time corrections could be applied when using GLAO.

\subsection{DEIMOS}

\begin{figure}
\begin{center}
\includegraphics[height=3in]{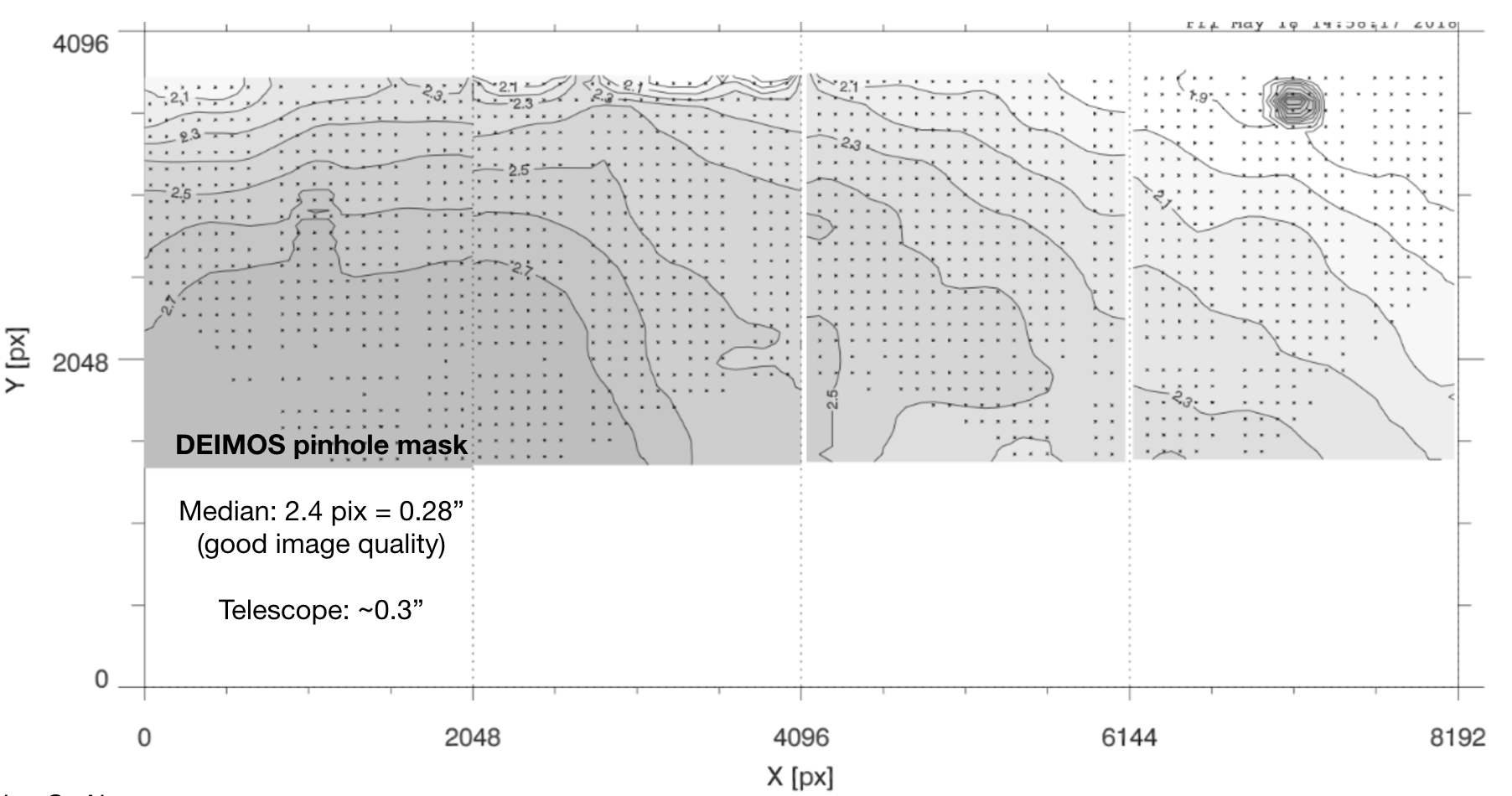}
\end{center}
\caption{\label{fig:deimos_iq}
DEIMOS image quality as measured by the FWHM of pinhole mask spots over the field of view. }
\end{figure}

We investigated whether ground-layer adaptive optics (GLAO) can benefit spectroscopy with the DEIMOS spectrograph on the Keck telescope. DEIMOS is a wide field multi-object spectrograph designed for long exposure extra-Galactic work, particularly observation of distant, early, galaxies. The DEIMOS science field is 16.7' $\times$ 5' with a pixel scale of 0.1185'' pixel$^{-1}$ for both the spectrograph and imager. This would be sufficient to Nyquist-sample a PSF that has been sharpened by GLAO. The intrinsic image quality within DEIMOS is has been characterized using pinhole mask observations (Figure \ref{fig:deimos_iq}) and is 0.26'' on average from the instrument. However, the telescope has an increasing astigmatism that increases with distance from the telescope's optical axis and would yield and additional 0.3'' (added in quadrature) on average over the DEIMOS field. Finally, DEIMOS has not been equipped with an atmospheric dispersion compensator. During long exposures, this could cause objects to shift outside the slit due to atmospheric refraction. 

\subsection{MOSFIRE}

MOSFIRE is a infrared imager and multi-object spectrograph located at the Cassegrain focus of the Keck I telescope. It has a 6.1'$\times$6.1' field of view and a plate scale of 0.18'', which is a bit coarse for sampling a 0.3''-0.4'' FWHM PSF from a GLAO feed. Ideally, MOSFIRE could be upgraded with a new detector (i.e. H4RG) with more and smaller pixels to preserve the field of view; but improve the sampling. The image quality inside MOSFIRE is excellent -- typically smaller than a single pixel \cite{McLean:2012}.  However, MOSFIRE lacks an atmospheric dispersion corrector, similar to DEIMOS, and would likely need one added to take full advantage of a GLAO feed. 

\section{Conclusions}
We study the feasibility of deploying a GLAO system at the W.~M.~Keck Observatory. A GLAO system at Keck would likely deliver images with a 2$\times$ better FWHM, resulting in improved point source sensitivity, better image resolution, and possibly higher spectral resolution through the use of narrower slits on GLAO-fed spectrographs. While Keck's existing multi-object spectrographs (LRIS, DEIMOS, MOSFIRE) have sufficient instrumental image quality to take advantage of a GLAO feed, they each have a shortcoming that requires further investigation before we can determine the final delivered image quality. First, LRIS suffers from flexure as it is a Cassegrain mounted instrument. The degree of flexure may be large enough to prevent long exposures with a narrow (0.3'') slit. Second, DEIMOS and MOSFIRE both lack an atmospheric dispersion compensator. Again, narrowing the slit without an ADC could cause objects to shift out of the slit at certain wavelengths at low elevations due to atmospheric refraction. Installation of an ADC for DEIMOS and MOSFIRE will be explored. Investigations are also underway to determine the feasibility of implementing laser guide star wavefront sensors in front of each of the instruments, and of implementing a 1.4 m diameter adaptive secondary mirror at Keck.

\acknowledgments 
We acknowledge funding from the W.~M.~Keck Observatory and the University of California Observatories to support the development of a GLAO concept for the Keck telescopes.    


\bibliography{report}   
\bibliographystyle{spiejour}   

\end{document}